\documentstyle{mn}
\input {psfig.sty}
\def \bj {b_{\rm J}}
\def \mb {M_{b_{\rm J}}}
\def \mbh {M_{\rm BH}}
\def \mdh {M_{\rm DH}}
\def \mgii {Mg\,{\sc ii}}
\def \civ {C\,{\sc iv}}
\def \lam {$\lambda$}
\def \Mpc {~h^{-1}~{\rm Mpc} }
\def \msun {\rm M_{\odot}}
\def \ledd {L_{\rm Edd}}
\def \hb {H$\beta$}

\title[Evolution of host mass vs. black hole mass]
      {The evolution of host mass and black hole mass in QSOs from the
      2dF QSO Redshift Survey}

\author[S. Fine et al.]
       {S. Fine$^{1,2}$\thanks{sfine@physics.usyd.edu.au},
       S.~M. Croom$^2$, L. Miller$^3$, A. Babic$^3$, D. Moore$^2$, 
       B. Brewer$^2$
\newauthor R.~G. Sharp$^2$, B.~J. Boyle$^4$, T. Shanks$^5$, R.~J. Smith$^6$,
       P.~J. Outram$^5$, N.~S. Loaring$^7$ \\
$^1$School of Physics, University of Sydney, NSW 2006, Australia \\
$^2$Anglo-Australian Observatory, PO Box 296, Epping, NSW 1710,
      Australia.\\
$^3$Department of Physics, Oxford University, Keble Road, Oxford, OX1
      3RH, UK.\\
$^4$Australia Telescope National Facility, PO Box 76, Epping, NSW
      1710, Australia.\\
$^5$Physics Department, University of Durham, South Road, Durham, DH1 3LE,
UK.\\
$^6$Astrophysics Research Institute, Liverpool John Moores University,
Twelve Quays House, Egerton Wharf, Birkenhead,  CH41 1lD, UK\\
$^7$Mullard Space Science Laboratory, Holmbury St. Mary, Dorking,
Surrey, RH5 6NT, UK
}

\begin{document}

\maketitle

\begin{abstract}
We investigate the relation between the mass of super-massive
black holes ($\mbh$) in QSOs and the mass of the dark matter halos
hosting them ($\mdh$).  We measure the widths of broad emission lines
(\mgii\ \lam2798, \civ\ \lam1549) from QSO composite spectra as a
function of redshift.  These widths are then used to determine virial
black hole mass estimates.

We compare our virial black hole mass estimates
to dark matter halo masses for QSO hosts derived
by Croom et al. (2005) based on measurements of QSO clustering.  This
enables us to trace the $\mbh-\mdh$ relation over the
redshift range $z=0.5$ to 2.5.
%The data are consistant with a constant
%relation between $\mbh$ and $\mdh$.
We calculate the mean
zero-point of the $\mbh-\mdh$ relation to be
$\mbh=10^{8.4\pm0.2}\,{\rm \msun}$ for an $\mdh=10^{12.5}\,{\rm 
  \msun}$.  These data are then compared with several models
connecting $\mbh$ and $\mdh$ as well as recent hydrodynamical
simulations of galaxy evolution. We note that the flux limited
nature of QSO samples
can cause a Malmquist-type bias in the measured zero-point of
the $\mbh-\mdh$ relation. The magnitude of this bias depends on the
scatter in the $\mbh-\mdh$ relation, and we reevaluate the zero-point
assuming three published values for this scatter.

We create a subsample of our data defined by a constant magnitude
interval around $L^{*}$ and
find $(1+z)^{3.3\pm1.3}$ evolution in $\mbh$ between $z\sim0.5$ and
$2.5$ for typical, $L^{*}$ QSOs. We also determine the Eddington
ratios ($L/\ledd$) for the
same subsample and find no significant evolution: $(1+z)^{-0.4\pm1.1}$.
Taken at face value, our data
suggest that a decrease in active black hole mass since $z\sim2.5$ is
the driving force behind luminosity evolution of typical, $L^{*}$,
optically selected AGN. However, we note that our data are also
consistent with a picture in which reductions in both black hole mass
and accretion rate contribute equally to luminosity evolution. In
addition we find these evolution results are strongly affected by the virial
black hole mass estimators used. Changes to the calibration of these
has a significant effect on the evolution results.

\end{abstract}


\section{Introduction} \label{sec:intro}

Over recent years it has become apparent that massive black holes
lie at the centre of the majority of local galaxies. In addition,
correlations between the mass of these central black holes ($\mbh$)
and host galaxy properties suggests that the evolution of
galaxies must be intimately related to the growth of their central
black holes.  Galaxy properties which display a correlation with
$\mbh$ include spheroid luminosity \cite{mag98}, spheroid mass
\cite{k+r95,fer02} and stellar velocity dispersion \cite{f+m00,geb00}.

In this paper we investigate the relationship between the mass of QSO
black holes and the mass of the dark matter halos that host them. Dark
matter halo masses derived from QSO clustering were calculated for
QSOs in the 2dF QSO Redshift Survey (2QZ; Croom et al. 2004) by Croom
et al. (2005). These were evaluated for QSOs in 10 redshift bins from
$z\sim0.5$ to $2.5$. We take the identical sample to that used by Croom
et al. (2005) and calculate average virial black hole masses for the
QSOs in each redshift bin. This is achieved by constructing a composite
spectrum for each redshift bin from all of the QSOs in that bin. The
average virial black hole masses are then estimated from the widths of
broad emission lines in the composite spectra.

In Section~\ref{sec:bhmass} we describe the virial black
hole mass estimators used in this paper, section~\ref{sec:d+a}
describes our data and analysis including composite making
and line width measurement.
In Section \ref{sec:dhmass} we briefly review the clustering
measurements which are used to derive dark halo host mass.  Our
results are presented in Section \ref{sec:res} and we present our
conclusions in Section \ref{sec:conc}.

Throughout this paper we assume a
flat $(\Omega_{\rm m},\Omega_{\Lambda})=(0.3,0.7)$,
$H_{0}=70\,{\rm km\,s}^{-1}\,{\rm Mpc}^{-1}$ cosmology.

\section{Black Hole Mass Estimates}\label{sec:bhmass}

At the basis of measuring QSO black hole masses is the virial theorem
such that the black hole mass $\mbh\approx
rv^{2}/G$, where $v$ is the
velocity dispersion of material gravitationally bound to the black
hole at a distance $r$ from it. Direct measurements of $r$ and $v$ have
been taken, for relatively nearby systems, in a variety of ways. Most
notably reverberation mapping programs have succeeded in accurately
measuring $\mbh$ for tens of local QSOs \cite{wpm99,kasp00,pet04}. This
technique requires careful long-term observations of the variability
in QSO spectra and can take years to produce results.
Therefore there has been considerable effort
put into finding quicker, simpler methods for estimating black hole 
masses which can be extended to higher redshifts.

One result of reverberation studies of nearby QSOs was the discovery
of a strong correlation between the radius of the H$\beta$ emitting
region around AGN, $r_{{\rm H}\beta}$, and the continuum luminosity at
5100\,{\AA} \cite{kasp00,kasp05}. $v_{{\rm H}\beta}$ can be found by the
relation $v_{{\rm H}\beta}=f\cdot {\rm FWHM}({\rm H}\beta)$ where
FWHM(H$\beta$) is the measured full width at half maximum of the
H$\beta$ spectral line, and $f$ is factor of order unity which depends
on the geometry of the broad line emitting region around the
AGN. Thus, using the width of the H$\beta$ line and the continuum
luminosity at 5100\,{\AA} single epoch estimates for $\mbh$ can
be made \cite{kasp00,vest06}.

In higher redshift QSOs the H$\beta$ line is redshifted out
of the optical spectrum. At these higher redshifts UV lines can
be used to try and estimate $\mbh$. Unfortunately, while the velocity
dispersions, $v_{\rm UV}$, can readily be measured there is no direct
way to measure the size of the of the emitting region $r_{\rm
  UV}$. Hence these estimators must be calibrated with H$\beta$
measurements of the same object. These calibrations have been performed
for the \mgii\ and \civ\ lines by McLure \& Jarvis (2002) and
Vestergaard (2002) respectively. The relation for \mgii\ was
then revised for luminous QSOs by McLure \& Dunlop (2004), and the
relation for \civ\ was revised for an updated cosmology by Vestergaard
\& Peterson (2006). The
resulting $\mbh$ estimators used throughout this paper are
\begin{equation} \label{equ_m+j}
\frac{\mbh}{{\rm M}_{\odot}}=3.2\left(\frac{\lambda
  L_{3000}}{10^{37}\,{\rm W}}\right)^{0.62}\left(\frac{{\rm FWHM(Mg\,{\sc
  ii})}}{{\rm km\,s}^{-1}}\right)^{2}
\end{equation}
for the \mgii\ line and
\begin{equation} \label{equ_ves}
\frac{\mbh}{{\rm M}_{\odot}}=4.6\left(\frac{\lambda
  L_{1350}}{10^{37}\,{\rm W}}\right)^{0.53}\left(\frac{{\rm FWHM(C\,{\sc
  iv})}}{{\rm km\,s}^{-1}}\right)^{2}
\end{equation}
for the \civ\ line. Here $\lambda L$ denotes the continuum
luminosities at the specified wavelength and FWHM() corresponds to the
measured full width at half maximum of the spectral line.

\section{Data and Analysis} \label{sec:d+a}

We are concerned with finding mean black hole masses for QSOs in the
redshift bins shown in table~\ref{tab:res}. Since these predominantly
cover redshift ranges where the H$\beta$ line is shifted off the end
of the visible spectrum we exploit the methods of McLure \& Jarvis
(2002) and Vestergaard \& Peterson (2006) for calculating these
masses. To this
end we require measurements of \mgii\ and \civ\ velocity
widths, as well as monochromatic continuum luminosities near these
lines (3000\,{\AA} and 1350\,{\AA} respectively). It should be noted
that both of these UV mass estimators exhibit considerable scatter.
It has been
suggested that this may be intrinsic, possibly due to geometric
considerations of the AGN \cite{m+d02,smit02}. However, in our
analysis we make use of composite spectra created by combining all of
the individual 2QZ spectra in the redshift bins ($\sim2000$ objects
per bin). Composite spectra have
several advantages. The high signal-to-noise they provide allows
high precision in the measurement of line widths, and combining many
spectra should average over intrinsic (e.g. geometric) variations.

In this section we briefly describe the 2QZ spectral data and our
method for constructing composite spectra from them. We then discuss
the process by which we measured the width of \mgii\ and \civ\ lines in
the composites, and finally how we calculated the monochromatic
luminosities $\lambda L_{3000}$ and $\lambda L_{1350}$.

\subsection{QSO spectra}

All of the data in this paper come from the 2QZ and are described in detail
elsewhere \cite{2qz12}. The sample contains $>23\,000$ spectra
of QSOs in the magnitude range $18.25<b_{j}<20.85$ observed with the
two-degree field instrument on the AAT.
Spectra cover the wavelength range
3700-7900\,{\AA} with a dispersion of 4.3\,{\AA}\,pixel$^{-1}$
and instrumental resolution of 9\,{\AA}. Spectra were typically
observed for 3300-3600\,s giving a median signal-to-noise ratio of 
$\sim5.0$\,pixel$^{-1}$.

%While in many spectra the poor signal to
%noise can hamper attempts to measure line withds, we can obtain an
%average line width for QSOs in a redshift bin by constructing a
%composite spectra.

\subsection{Composite spectra}

The use of composite spectra is not a new technique (e.g. Francis et
al. 1991; Vanden Berk et al. 2001) and a full description of our
method for creating composites is given by Croom et al. (2002). Here
it is worth noting that 2QZ spectra are not flux
calibrated. Therefore, to make the composites the individual spectra 
had to be normalised by fitting a polynomial continuum to regions
without emission features.  We then divide by this fitted continuum to
uniformly normalise the spectra before combining.  In doing so all
information on continuum slope and normalisation is lost while the
emission features remain intact. The normalised spectra were then
shifted to the rest frame and the composites were constructed as the
median of all contributing QSOs in each pixel (of width 1\,\AA).
Errors were determined by taking the 68 per cent semi-interquartile
range of individual QSO pixel values.

\subsection{Measurement of Line Widths} \label{ssec_lw}

\begin{figure*}
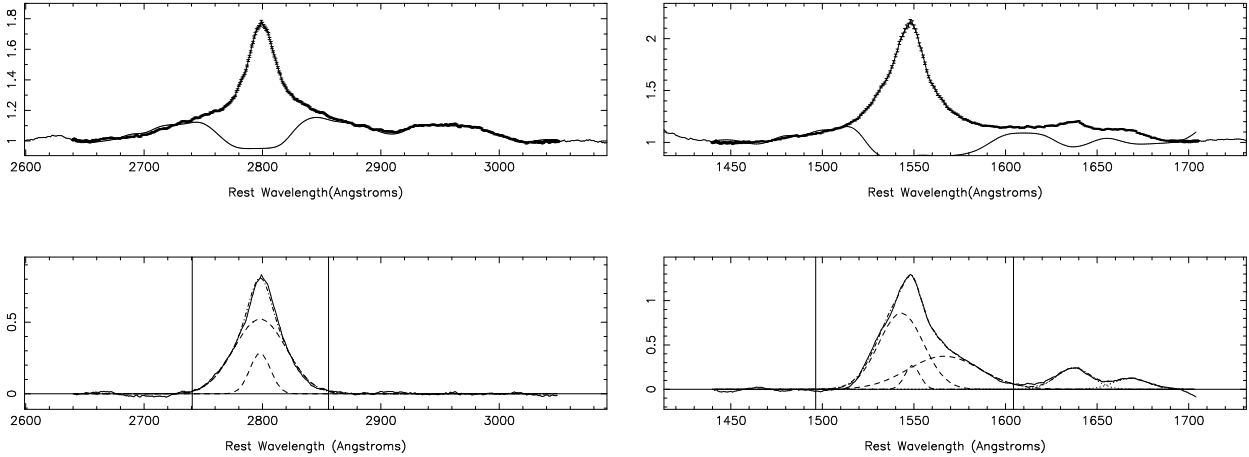

\centering
\centerline{\psfig{file=composite_z092to113_Mall_Mgii.ps,width=8.0cm,angle=-90}\hspace{0.5cm}\psfig{file=composite_z202to225_Mall_Civ.ps,width=8.0cm,angle=-90}}
%\centerline{
%\psfig{file=composite_z202to225_Mall_Civ.ps,width=8.0cm,angle=-90}}
\caption{Illustrations of the line fitting procedures for \mgii\
  (left) and \civ\ (right) in two of the composite spectra, note we
  observe very little variation in the line profiles between
  composites. In each case the top panel shows
  the initial composite spectrum and the smoothed iron template fitted to
  it. The lower panels show the spectrum after the iron emission was
  subtracted from it. Also shown in the bottom panels are the
  Gaussians used to model the line (dashed), as well as Gaussians used
  to fit other local emission features (dotted) and the sum of these
  for comparison with the composite (dash-dot; which follows
  closely the solid line). Vertical lines in the bottom panels
  indicate the regions outside of which the iron template
  was fit to our data. 68\% interquartile errors for
  each point in the composite spectra are shown in the upper panels
  (these are hardly visible)
  but are omitted from the lower for clarity. For each spectrum the
  y axis denotes $F_{\lambda}$ in arbitrary units of flux density.}
\label{fig_line-fits}
\end{figure*}

Before we measured emission line widths from our composite spectra, we
had to correct for QSO iron emission. This correction was
performed iteratively. A smoothed template
of QSO iron emission \cite{ves01} and linear continuum were fitted to
the data in regions either side of the
emission line in question. The iron template and continuum were then
subtracted from the data, and a
single Gaussian profile was
fitted to the remaining line. In the case of C\,{\sc iv} two other 
Gaussians were also fitted to the He\,{\sc ii}/O\,{\sc ii}]
feature just redwards of the line. The iron template and continuum
were then fitted to
the spectrum again, this time excluding any data within
the primary emission line region (defined as $\pm3\sigma$ of the
Gaussian fit
to the line) and in the case of C\,{\sc iv} with the two
other Gaussians (defining the He\,{\sc ii}/O\,{\sc ii}] flux)
subtracted from the data. This process was repeated
until the width of successive Gaussian fits to the line differed by less than
half their associated error. It is worth noting here that the Gaussian
we fit to the primary emission line is not used to measure its width
but only as a mask. Neither the
\mgii\ nor the \civ\ line are well described by a single Gaussian,
instead this allows us to define the parts of the spectrum
unaffected by the emission line to use in our subsequent iron/continuum
fit. Using this method we found that we could
accurately remove the local emission features around both of these
emission lines
without making preliminary assumptions as to their width
(see Fig.~\ref{fig_line-fits}).

Once the iron emission had been subtracted from the spectra it was
possible to measure the width of the lines. Our method for
measuring line widths is similar to that used by Wang, Lu \&
Zhou~(1998) in that we model each line's profile with a set of
Gaussian components. We then take the FWHM of this model as our line
width. We found this process to be more robust than reading the FWHM
directly from the data which can be significantly affected by a single
pixel, and a better indicator of the true FWHM than attempting to
impose a profile on the line such as a single Gaussian or Lorentzian.

Since the \mgii\ line is symmetric we can accurately model
its profile with two Gaussians. During the fitting process we tie the central
wavelength of the Gaussians together but all other parameters (five in
total: the central wavelength of the line and, the width and amplitude
of each Gaussian component fit to the line) were left free. The
asymmetry of the \civ\ line requires that we
use three Gaussians to accurately model its profile with all
nine parameters left free in the fitting process. In each case
the multi-Gaussian models were fit to the data using
the {\sc mrqmin} routine \cite{press89}.
At each step in the above fitting we take into account the
propagation of errors through the process, and we reevaluate the error at each
pixel taking into account uncertainty in the iron and continuum fits.
The errors on the FWHM measurements could then be calculated
analytically from the covariance matrix of the multi-Gaussian fit.
Fig.~\ref{fig_line-fits}
illustrates the line fitting process
for both \mgii\ and a \civ. The multiple Gaussian
components (two for \mgii\ and three for \civ) provide accurate
models for the emission lines.

After each line width had been measured we corrected for the
resolution of the spectrograph (9\,\AA\ rest frame) by subtracting it
in quadrature from the measured line width.

In calibrating equations~\ref{equ_m+j} and~\ref{equ_ves} McLure \&
Dunlop and Vestergaard \& Peterson define their FWHMs in different ways. Since
$r$ in the virial equation
defines the radius of the broad-line region surrounding AGN McLure \&
Dunlop correct for narrow line \mgii\ emission, and
take the FWHM of only the the broad component of the line to calibrate
equation~\ref{equ_m+j}. 
However, in their analysis of broad UV lines in QSO spectra Wills et
al.~(1993) found no evidence for a narrow contribution to these
lines. Therefore, Vestergaard \& Peterson (2006) measure the FWHM of
the entire \civ\ line when calibrating equation~\ref{equ_ves}.
%Vestergaard argues that in measuring the velocity width from
%a single epoch QSO spectra, one is trying to reproduce the velocity
%widths measured from rms QSO spectra derived from reverberation
%mapping. She shows that some lines in the rms spectra exhibit strong
%narrow line cores and thus concludes that this narrow line emission
%must originate from the broad line region.

McLure \& Dunlop~(2004) model the \mgii\ line with a broad and narrow
component following a similar procedure to that which we have outlined above.
They then record the width of the broader component of
the line to use in their calculations. However, in their fitting they
impose the additional conditions that the velocity width of the
narrower component of the line be $<2000$\,km\,s$^{-1}$, and that the
equivalent width of the narrower component be less than one third the
equivalent width of the broad component. We do not apply these
constraints in our fitting process and find that, while the narrower
components all have equivalent width less than 1/3 that of the broad,
in all but one composite the width of the narrower \mgii\ component is
$>2000$\,km\,s$^{-1}$. If we were to add this additional constraint to
the fitting procedure it would degrade the quality of the line fits,
although we find this effect is small. Wills et al.~(1993) find no
evidence for a narrow line component in the \mgii\ line in QSO spectra
and we find no reason to subtract off this `narrower' contribution to
the line. We also note that the use of Gaussians in our fitting
process is somewhat arbitrary and hence it is difficult to assign a physical
meaning to each of the components individually. We thus use the FWHM
of the whole line in our calculations for \mgii\ (this may introduce
systematic errors to our calculations, see section~\ref{sec:err}).

Note that the asymmetry of the
\civ\ line (bottom-right panel Fig.~\ref{fig_line-fits}) is of concern
to this analysis. Other investigations of UV QSO spectra have shown
that the \civ\ line is often found to be blueshifted with respect to
lower ionisation lines such as \mgii\ (e.g. Marziani et al. 1996), and
Richards et
al.~(2002) suggested that this may imply the \civ\ emitting region is
outflowing from the AGN and hence not be fully virialised. However, our
\civ\ line profiles do not resemble those in
composite spectra constructed by Richards et al. for QSOs with
blueshifted \civ. The asymmetry we observe in the \civ\ line could be
caused by a range of possible physical processes including emission
and/or absorption by non-virialised gas, however, a full discussion of
this is beyond the scope of this investigation.
%Note that the asymmetry of the
%\civ\ line (bottom-right panel Fig.~\ref{fig_line-fits}) is a concern
%as this may imply that the gas emitting this
%line is not fully virialised. Other evidence for this comes from the
%well documented
%blueshift of the \civ\ line with respect to lower ionisation
%lines in QSO spectra such as \mgii\ 
%\cite{mar96,rich02}. This may indicate that instead of being
%virialised the \civ\ emitting gas is associated with an outflow from
%the AGN. If this is the case our measured \civ\ line widths cease to
%be an accurate indicator of the virial velocity dispersion and we
%cannot use equation~\ref{equ_ves} to find black hole masses.

On the
other hand, we note that the original \civ\ line (top-right
panel Fig.~\ref{fig_line-fits}) shows no obvious asymmetry. The
asymmetry then becomes pronounced after the iron template has been
subtracted from the data. Therefore, it is possible that the asymmetry
we observe may not be caused by any physical process in the
QSOs, but by bad iron subtraction. Vestergaard \& Wilkes (2001) make note
when creating the template that iron emission in the vicinity of the
\civ\ line can be difficult to isolate. The iron
template is based on spectra of the Seyfert~I galaxy
I~Zwicky~1, which shows considerable Si\,{\sc ii} emission 
bluewards of the \civ\ line. The \civ\ line itself is unusually weak,
making it
difficult to deblend the carbon, silicon and iron
emission. Vestergaard \& Wilkes (2001) find no emission directly redwards of
the \civ\ line, and hence the iron template is asymmetric about this
line. It is therefore unclear whether the asymmetry of the \civ\ line
after the iron emission has been subtracted is real. We find no
evidence of Si\,{\sc ii} emission in any of our composites, and it
may be that this iron template is simply not applicable in this case.

We thus took note of the effect of the iron subtraction on our
data. We repeated our analysis but instead of fitting an iron
template along with a continuum to the data, we simply used a linear
continuum fit between the approximate extremities of the \civ\ line
(1500\,{\AA} and 1600\,\AA). We found this gave us FWHMs which were
consistently smaller by a factor $\sim1.3$, however, the symmetry of
the line remained intact in this procedure. In the analysis that
follows we only use data measured after we had corrected for iron
emission. Equation~\ref{equ_ves} was calibrated with line widths
measured after correcting for iron emission, and we find that the
template does accurately fit our composite spectra outside the immediate \civ\
line region.

\subsection{Luminosity Measurements}

The absolute magnitudes quoted in this paper ($\mb$)
come from photographic observations in the $\bj$ band with the UK
Schmidt. These are corrected for extinction by Galactic dust
\cite{sfd98}, and $K$-corrected using the values provided by
Cristiani~\&~Vio~(1990). To then calculate continuum luminosities at
the wavelengths desired we
made use of the Sloan Digital Sky Survey (SDSS) QSO composite
spectrum \cite{van01}. We first calculated the mean redshift and $\bj$
luminosity of the QSOs in our composite which contributed to the spectral
line we were analysing. Then the SDSS composite was redshifted
to $\overline{z}$, and normalised to the measured luminosity in the $\bj$
band by convolving with the UKST $\bj$ response function.  Then it was
possible to read off values for $L_{1350}$ and $L_{3000}$.

The magnitudes listed in the 2QZ are generally accurate to $0.1-0.2$
magnitudes. Correcting for dust and K-correcting could introduce
significant uncertainty into the values for $\mb$. However, once
averaged over all $\sim2000$ QSOs in a redshift bin we assume the 
error on $\overline{M}_{b_{J}}$ will be negligible. On the other hand the
extrapolation to $L_{1350}$ and $L_{3000}$ could be a significant
source of error as we discuss below.

%Extrapolating luminosities in this manner can often be a large source
%of error in calculations especially when performed on single
%objects. In our case, however, we are calculating average luminosities
%for the $\sim2000$ QSOs in each redshift bin. Thus
%using the flux calibrated SDSS composite, itself constructed from
%$>2000$ QSO spectra in the redshift range we are investigating, should
%introduce only small errors into the luminosity
%calculations. Considering the large ($>0.3$\,dex) scatter
%associated with equations~\ref{equ_m+j} and~\ref{equ_ves}, we assume
%that errors in the luminosity calculations will be negligible and we
%make no attempt to quantify them here.

\subsection{Errors} \label{sec:err}

There are three possible sources of error to our calculations for
$\mbh$. These are errors in the luminosity values, errors in the line
width measurements and intrinsic scatter associated with
equations~\ref{equ_m+j} and~\ref{equ_ves}.

Equations~\ref{equ_m+j} and~\ref{equ_ves} are quoted as having errors
of 0.33\,dex and 0.36\,dex respectively \cite{m+d04,vest06}, and as we
shall see these large scatters dominate the random error in
$\mbh$. It is worth noting here that since the quoted errors are rms
values and because we are dealing with large
numbers of objects ($\sim2000$
QSOs per redshift bin), it could be argued that the errors
should be reduced by the according factor
($\sim1/\sqrt{2000}$). However, the large error in the
virial mass estimators is to a large extent due to the limited number
of AGN with reliable mass
estimates from reverberation mapping. Since the calibrations are
based only on a few tens of objects, we do not believe it to be
prudent to be reducing the errors in our calculations because of the
large number of objects in our dataset.

The high signal to noise of composite spectra mean that very precise
line widths can be measured. We calculate errors on these widths
analytically from the covariance matrix of the
fitting parameters described in section~\ref{ssec_lw}, and find them to
be negligible (see table~\ref{tab:res}). However, the process by which we
measure line widths could introduce significant systematic errors to our
calculations. One source of uncertainty is the nature of the iron template we
used and how it was fit to our data. We note above that the asymmetry
we observe in the \civ\ line may be a product of the iron template we
are using, and we cannot be sure that using this iron template doesn't
introduce errors into the line width measurements. We find that if we
simply fit a linear continuum around the \civ\ line our results are
reduced by $\sim1.3$. We therefore gauge that any problems with the
iron template will have an effect no greater than this on our line
width measurements, or a factor of $\sim1.7$ in our virial
black hole masses.

Another source of uncertainty in our line width determinations
is narrow line emission. None of the spectral lines analysed show
clear signs of narrow line emission. Their profiles are smooth and do
not exhibit the inflection characteristic of narrow line emission
superimposed on a broad line. Indeed other investigations of the UV lines in
QSO spectra have found no evidence that they contain narrow line cores
\cite{wills93}. However, as stated above McLure \&
Dunlop only consider the broader component of the \mgii\ line when
calibrating equation~\ref{equ_m+j} while we measure the FWHM of the
whole line. In doing so we may
introduce systematics to our calculations. We consider
the possible effect of this by recording the width of the
broader component of the modelled \mgii\ fit. We find that line widths
measured this way are
$\sim1.5$ times larger than those measured for the whole line which
translates to roughly a factor of 2 in $\mbh$.

Extrapolating luminosities in the manner described above can often be
a large source of error in calculations, in particular when performed
on single objects. In our case, however, we are calculating average
luminosities for the $\sim2000$ QSOs in each redshift bin. Thus
using the flux calibrated SDSS composite, itself constructed from
$>2000$ QSO spectra in the redshift range we are investigating, should
introduce only small errors into the luminosity
calculations. The SDSS median QSO composite has a continuum described
by $F_{\lambda}\propto\lambda^{\alpha}$ with $\alpha=-1.54$ for
$\lambda<5000$\,{\AA} and $\alpha=-0.42$ for
$\lambda>5000$\,{\AA}. Power laws with $-2<\alpha<-1.5$, are used
throughout the literature to make extrapolations similar to those
we make. Therefore we also evaluate the monochromatic luminosities
assuming these power laws and define the error on our luminosity
values to be half their difference.

We calculate the errors on the virial mass estimates
taking into account the scatter
in the virial mass 
estimators as well as errors in our luminosity calculations and line
width measurements. However, the the final error on the mass estimate is
completely dominated by the uncertainty in the mass estimator.
We do not attempt to account
for possible sources of uncertainty from poor iron subtraction or narrow
line emission in the tabulated errors.

\section{Dark Halo Mass}\label{sec:dhmass}

Croom et al. (2005) used measurements of the clustering of QSOs from
the 2QZ to infer the mass of dark matter haloes that they inhabit.
They divide the 2QZ sample into 10 redshift intervals (see table
\ref{tab:res}) each containing $\sim2000$ QSOs and measure the
two-point correlation function (accounting for redshift-space
effects) on $<20\Mpc$ scales.  They then compare this to the evolution
of mass clustering in a WMAP/2dF cosmology \cite{wmap,2dfgrspk02} to
determine the bias of QSOs as a function of redshift. Note that Croom
et al. (2005) use a slightly different cosmology in their
calculations: $(\Omega_{\rm m},\Omega_{\Lambda})=(0.27,0.73)$ and
$H_{0}=73\,{\rm km\,s}^{-1}\,{\rm Mpc}^{-1}$, although we do not
expect these slight differences to have a significant affect on our
calculations.

Finally they
use the formalism developed by Mo \& White (1996) to relate bias and
dark matter halo mass, specifically using the relation for ellipsoidal
collapse given by Sheth, Mo \& Tormen (2001).  This results in the
finding that 2QZ QSO hosts have approximately the same dark matter
halo mass as a function of redshift, with
$\mdh=(3.0\pm1.6)\times10^{12}h^{-1}\,{\rm \msun}$.  Croom et al. (2005) then
use a range of partly theoretical relationships between $\mbh$ and
$\mdh$ to derive black hole masses for these QSOs (see Eqs 22-26 in
Croom et al 2005).  In all cases the black hole masses are seen to
decrease toward lower redshift.  This so called {\it cosmic downsizing}
(see also Barger et al. 2005; Heckman et al. 2004) appears to be
driving QSO luminosity evolution at $z<2.5$.  However, in order to
demonstrate this more conclusively, we have undertaken the analysis in
the present paper to determine more directly the black hole masses of
2QZ QSOs (building on the previous work of Corbett et al. 2003). 

\section{Results} \label{sec:res}

\begin{figure}
\centering
\centerline{\psfig{file=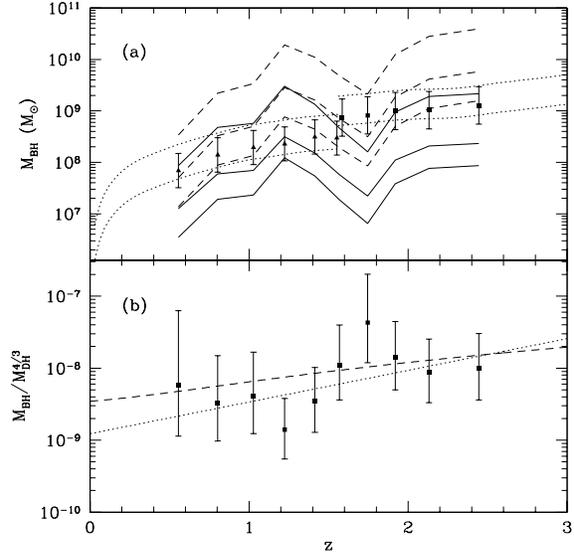,width=8.0cm}}
\caption{(\emph{a}) The trend in virial $M_{\rm BH}$ estimates with
  redshift. The
  difference between the Mg\,{\sc ii} (triangles) and C\,{\sc iv}
  (squares) results is well illustrated here. We also show
  approximate limits on our black hole mass estimates as imposed by
  the flux limits of the 2QZ assuming an average line width (dotted lines).
  In addition $M_{\rm BH}(\mdh)$ estimates calculated assuming various
  dark matter halo density profiles with no
  evolution in the $\mbh-\mdh$ relation (solid lines) and evolution
  $\propto(1+z)^{2.5}$ (dashed lines) are shown. For each model the
  calculations assuming an
  isothermal density profile give the lowest estimates for $M_{\rm
  BH}$, the intermediate estimates are given by an NFW profile, and the
  Seljak model gives the highest. Note that we do not present the
  errors on these points for clarity.
  (\emph{b})
  We show our values for $\mbh/\mdh^{4/3}$ (squares) compared with the 
  predictions of Robertson~(2005) (dashed line). We also show our best
  fit to the data (dotted line).}
\label{fig_z-bhm}
\end{figure}

Table~\ref{tab:res} displays the results of our
analysis and in Fig.\ref{fig_z-bhm}a we show the virial black hole
mass estimates as a function of redshift. A clear `step' is visible
where we switch from \mgii\ estimates to \civ. In addition, for one
redshift bin we have both \mgii\ and \civ\ present in the composite
spectra, and for this bin the calculated values of $\mbh$ differ
by just over $1\sigma$. This disagreement is only marginally
significant but does raise questions over the calibrations of
equations~\ref{equ_m+j} and \ref{equ_ves}. However, since these two
relations are not inter-calibrated we should not be surprised
that we find some discrepancy. We also note that the
magnitude of this offset is consistent with the systematic errors
discussed in section~\ref{sec:err}.
Hereafter we use the weighted mean of the two mass
estimates for the redshift bin $z\in(1.50,1.66)$:
Log$(\frac{\mbh}{\msun})=8.7\pm0.35$.

We note here that the \hb\ line is present in the composite spectrum
for redshift bin $z\in(0.30,0.68)$. We could therefore obtain a virial
black hole mass estimate for this bin from the \hb\ line following a
calibration by
e.g. Vestergaard \& Peterson (2006). Comparing this with
the \mgii\ mass estimate for the same redshift bin could help tie
the calibrations together, and potentially clarify why we observe the
difference between the \mgii\ and \civ\ estimators. However,
we found that procedures for measuring the width of the \hb\ line
were not well defined in the literature. In particular other authors
have found evidence for a `very broad component' to the \hb\ line
(e.g. Marziani 2003) which could have a large effect on our
measurements. In addition, because of the nature of composite spectra
the same set of individual QSO spectra do not contribute to the
\hb\ and \mgii\ lines in a single composite (note this is equally true
for the \civ\ and \mgii\ lines in composite $z\in(1.50,1.66)$). Due to
the above
considerations, and because we would only obtain a single point to compare
with our \mgii\ measurements, we do not make an estimate of the virial
black hole mass for the first redshift bin from the \hb\ line.

It is interesting to note that the line widths in table~\ref{tab:res}
hardly vary between the composites. Given that
this is the case any variation observed in $\mbh$ must be due to the
luminosity. Thus the flux limits of the 2QZ $(18.25<\bj<20.85)$ will
effectively impose limits on the masses we have calculated.
We can estimate these limits by calculating
the black hole mass as a function of redshift, given a source with
an average line width and an apparent magnitude at the upper and
lower limits of the survey.
We show these limits in Fig.\ref{fig_z-bhm}a and it appears that these
confine our average black hole masses to only a very small
range of possible values at any given redshift.
%However, it is worth noting that these are retrospective
%limits, and are only defined because we have found such little variation
%in line widths between our spectra. In principal there could have been
%more variation in the line widths and in this case they would not apply.
It is worth noting that the upper flux limit of the 2QZ does not
have a tremendous effect on our calculations. Croom et al. (2004)
extended the 2QZ to a $\bj$ of 16 in the 6dF QSO Redshift survey (6QZ)
and found an extra $\sim320$ QSOs in roughly half the area of sky
as surveyed in the 2QZ. Considering the comparatively small numbers of these
bright QSOs, we do not believe they could contribute significantly to
our composite spectra. The lower flux limit, however, does clearly
affect our results. Thus we do not present Fig.\ref{fig_z-bhm}a as evidence
for evolution in black hole mass for the global QSO
population. Instead this shows average virial black hole mass
estimates for QSOs in the 2QZ between redshift 2.5 and 0.5.

\begin{table*}
\begin{center}
\caption{Each line represents measurements taken from one spectral
  line (Note that the composite spectra for redshift bin
  $z\in(1.50,1.66)$ had both the \civ\ and \mgii\ lines visible).
  For each line we give the average redshift and absolute $\bj$
  magnitude of QSOs contributing to the composite spectra at that line,
  along with the absolute magnitude of the break in the QSO luminosity
  function at that redshift $\mb^{*}$ (Assuming the polynomial evolution
  model of Croom et al. 2004). Note that the values for $\overline{z}$
  and $\overline{M}_{\bj}$ vary at each point in a single composite
  spectra because at each point there will be a different group of
  spectra contributing to the composite. We also present the measured FWHMs for
  the lines and monochromatic luminosities near them
  3000\,{\AA} and 1350\,{\AA} for \mgii\ and \civ\
  respectively). We give the black hole mass calculated from the line
  and derived Eddington ratios $(L/\ledd)$ and finally the
  dark halo masses calculated by Croom et al.~(2005).}
\label{tab:res}
\setlength{\tabcolsep}{3pt}
\begin{tabular}{ccccccccccccc}
\hline \hline
$z$ interval & $\overline{z}$ & $\overline{M}_{\bj}$ &
$\mb^{*}$ & Spectral Line & FWHM ({km\,s$^{-1}$}) & Log($\frac{\lambda L}{\rm W}$) &
Log($\frac{\mbh}{\msun}$) & Log($L/\ledd$) & $\frac{\mdh}{\times10^{12}\msun}$ \\
\hline
0.30,0.68 & 0.556 & --22.28 & --23.30 & \mgii & $3546\pm19$ & $37.379\pm0.001$ & $7.8\pm0.33$ & --$0.71\pm0.33$ & $1.15^{+2.18}_{-0.94}$ \\
0.68,0.92 & 0.803 & --23.26 & --23.91 & \mgii & $3875\pm14$ & $37.747\pm0.017$ & $8.1\pm0.33$ & --$0.64\pm0.33$ & $2.94^{+3.07}_{-1.83}$ \\
0.92,1.13 & 1.028 & --23.85 & --24.39 & \mgii & $3878\pm15$ & $37.977\pm0.030$ & $8.3\pm0.33$ & --$0.56\pm0.33$ & $3.25^{+3.14}_{-1.93}$ \\
1.13,1.32 & 1.224 & --24.26 & --24.75 & \mgii & $3783\pm16$ & $38.127\pm0.040$ & $8.4\pm0.33$ & --$0.48\pm0.33$ & $8.11^{+4.08}_{-3.11}$ \\
1.32,1.50 & 1.414 & --24.57 & --25.04 & \mgii & $4104\pm18$ & $38.229\pm0.049$ & $8.5\pm0.33$ & --$0.50\pm0.33$ & $5.20^{+3.15}_{-2.28}$ \\
1.50,1.66 & 1.552 & --24.75 & --25.22 & \mgii & $3889\pm31$ & $38.272\pm0.055$ & $8.5\pm0.33$ & --$0.41\pm0.33$ & $2.89^{+2.27}_{-1.51}$ \\
1.50,1.66 & 1.585 & --24.80 & --25.26 & \civ  & $5438\pm58$ & $38.394\pm0.031$ & $8.9\pm0.36$ & --$0.79\pm0.36$ & $2.89^{+2.27}_{-1.51}$ \\
1.66,1.83 & 1.746 & --25.03 & --25.42 & \civ  & $5444\pm44$ & $38.472\pm0.024$ & $8.9\pm0.36$ & --$0.75\pm0.36$ & $1.62^{+1.75}_{-1.01}$ \\
1.83,2.02 & 1.919 & --25.25 & --25.56 & \civ  & $5687\pm40$ & $38.564\pm0.017$ & $9.0\pm0.36$ & --$0.75\pm0.36$ & $4.30^{+2.61}_{-1.89}$ \\
2.02,2.25 & 2.132 & --25.46 & --25.67 & \civ  & $5629\pm39$ & $38.609\pm0.010$ & $9.0\pm0.36$ & --$0.69\pm0.36$ & $6.28^{+3.10}_{-2.37}$ \\
2.25,2.90 & 2.445 & --25.83 & --25.71 & \civ  & $5707\pm39$ & $38.763\pm0.001$ & $9.1\pm0.36$ & --$0.64\pm0.36$ & $6.73^{+3.77}_{-2.80}$ \\
\hline \hline
\end{tabular}
\end{center}
\end{table*}

\subsection{The $\mbh-\mdh$ relation}

Correlations between black hole mass and galactic properties locally imply
that massive black holes grow in parallel with the galaxies (and
presumably dark matter halos) in which they reside.  Thus we expect
$\mbh$ and $\mdh$ to be related. Ferrarese~(2002)
proposed three such relations based the local $\mbh-\sigma$ relation
and different models for the dark matter halo density profile. In brief
the first assumes an isothermal dark matter profile, the second
assumes an NFW profile \cite{nfw97}, and the third assumes a profile
based on the weak lensing results of Seljak~(2002). In each of these
models the dark halo mass is calculated from an estimate for the
halo virial velocity $v_{\rm vir}$ which
cannot be directly measured for dark matter
halos. Hence Ferrarese extrapolates from the galaxy's circular velocity,
$v_{\rm c}$, to  $v_{\rm vir}$ using the above density profiles.
This is a large extrapolation and the source
of the offset between the models. We combine the models
with two assumptions concerning the evolution of this relation (see
Wyithe \& Loeb~2005; WL05), namely that $\mbh-\mdh\propto(1+z)^{2.5}$; dashed
lines in Fig.\ref{fig_z-bhm}a) and $\mbh-\mdh$ is constant with $z$ (solid
lines in Fig.\ref{fig_z-bhm}a).

Table~\ref{tab_bh-models} summarises
the results of $\chi^{2}$ analysis comparing the black hole
masses derived from the dark halo mass, $\mbh(\mdh)$, to our virial
black hole masses. That is, a comparison between our data points and
the solid/dashed lines in Fig.\ref{fig_z-bhm}a allowing for the
errors on the estimate of  $\mbh(\mdh)$ (which, for clarity, are not
shown in Fig. \ref{fig_z-bhm}a).  The comparison was done in log-space
and (for non-symmetric errors) took the error on $\mbh(\mdh)$ in the
direction of the virial $\mbh$ estimates.
The errors are not normally distributed, and hence our results here
are a guide rather than being statistically robust.  However, it
appears that the `$\mbh-\mdh$ constant' model with an isothermal
profile is rejected by the data, while the others are reasonably
acceptable (i.e. cannot be rejected at the 85\% level).

\begin{table}
\begin{center}
\caption{Comparisons of $\chi^{2}$ statistics for the different models
  shown in Fig.~\ref{fig_z-bhm}a.  Note that since errors on $\mdh$
  are not normally distributed this $\chi^{2}$ analysis is not
  statistically robust. We present it here as a guide to how well each
  model matches our data.}
\label{tab_bh-models}
\begin{tabular}{lccc}
\hline \hline
Model & Assumed dark matter & $\chi^{2}$ & P($>\chi^{2}$) \\
   & halo density profile & & \\
\hline
$\mbh-\mdh$          & Isothermal & 24.1 & 0.01 \\
  const.             & NFW        & 14.1 & 0.17 \\
                     & Seljak     & 5.0  & 0.89 \\
$\mbh-\mdh$          & Isothermal & 5.4  & 0.86 \\
$\propto(1+z)^{2.5}$ & NFW        & 6.7  & 0.75 \\
                     & Seljak     & 13.4 & 0.20 \\
\hline \hline
\end{tabular}
\end{center}
\end{table}

%%% SMC: don't know where this 4.6 came from:
%We also note that the $\mbh$ evolution in the case of
%$\mbh-\sigma$ constant $\sim(1+z)^{4.6}$ is inconsistent with our
%observations emphasising that the Ferrarese-Seljak model best
%describes our results.

Recent hydrodynamical simulations of galaxy mergers have shown that
the local $\mbh-\sigma$ relation, along with a startling array of
other QSO features, can be reproduced under the
condition that QSO energy feedback self-regulates the growth of
massive black holes (eg: Di Matteo et al. 2005; Robertson et al. 2005;
Hopkins et al. 2005). These simulations predict an $\mbh-\sigma$
relation of the form
\begin{equation}
{\rm log}\left(\frac{\mbh}{\rm \msun}\right)\approx 8.1 + 4.0\,{\rm
  log}\left(\frac{\sigma}{200\,{\rm km\,s^{-1}}}\right) - 0.19\,{\rm
  log}(1+z).
\label{eq:robertson}
\end{equation}
We take $\sigma \approx V_{\rm vir}$ (see Fig.~3 of Di Matteo et
al. 2005) where $V_{\rm vir}$ is the virial velocity in the
simulations and is directly related to the total galaxy mass by
$\mdh \approx M_{\rm vir}=V_{\rm vir}^{3}/10\,G\,H(z)$. Assuming the
cosmological parameters specified in section~\ref{sec:intro} this gives us
a redshift dependent relationship between $\mbh$ and $\mdh$
characterised by
{\setlength\arraycolsep{2pt}
\begin{eqnarray}
{\rm log}\left(\frac{\mbh}{\rm \msun}\right) & \approx & - 8.5 +
  \frac{4}{3}{\rm log}\left(\frac{\mdh}{\rm \msun}\right) \nonumber\\
  & & \hspace{0.5cm} - {\rm
  log}\left(\frac{(0.7+0.3(1+z)^{3})^{\frac{2}{3}}}{(1+z)^{0.19}}\right).
\label{eq:rob2}
\end{eqnarray}}
We show their predictions
for $\mbh / \mdh^{4/3}$ as a function of redshift in
Fig.~\ref{fig_z-bhm}b along with our calculated values.
We find good agreement between the simulation predictions
and our values.  Our best fit to the data is also shown which follows
$\mbh / \mdh^{4/3}\propto(1+z)^{2.5\pm1.8}$.

\begin{figure}
\centering
\centerline{\psfig{file=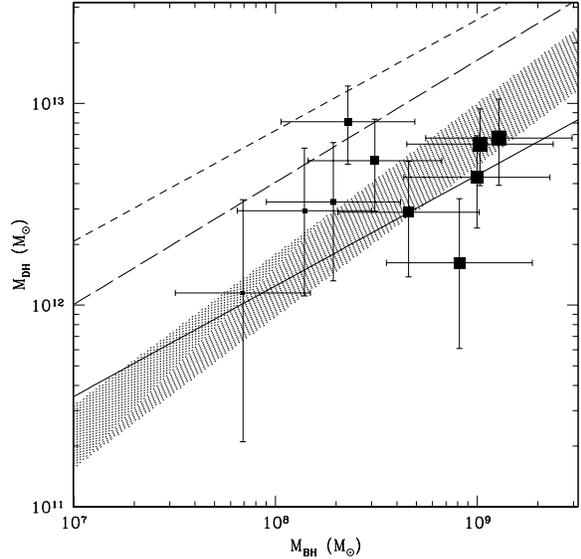,width=8.0cm}}
\caption{Shows our values for $\mbh$ plotted against $\mdh$ found by
  Croom et al.~(2005). The size of the points are scaled by redshift (larger
  points for higher $z$). The $\mbh-\mdh$ relation of
  equation~\ref{eq:rob2} is also shown as the shaded region in
  the plot, with the relation at $z=0$ defining the top of the region
  and $z=2.5$ defining the bottom. The $z=0$ $\mbh-\mdh$ relations of
  Ferrarese~(2002) are also shown (heavy lines) for the isothermal
  (short dash), NFW (long dash) and Seljak (solid) density profiles.} 
\label{fig_bhm-dhm}
\end{figure}

Fig.~\ref{fig_bhm-dhm} shows the relation between $\mbh$ and
$\mdh$. We observe a weak correlation between these values
significant at only the 85 per cent level (applying a Spearman rank
test). The weakness of this correlation is due in part to the limited
dynamic range of our averaged mass estimates.
We also plot
Eq.~\ref{eq:rob2} between $z=0.5$ and 2.5 as the
shaded region in the Figure, and the $z=0$ relations 
of Ferrarese~(2002).  We see that the relation of  Robertson et
al. is in excellent agreement with  our data.  It is worth noting that
that the normalisation of the simulated curve (in both
Fig.~\ref{fig_z-bhm}b and~\ref{fig_bhm-dhm}) comes only from the QSO
luminosity function. It is therefore encouraging that the simulation's
predictions for the $\mbh-\mdh$ relation match our measurements so
well.  The Ferrarese relation
assuming a Seljak profile is also in good agreement with the data.
However, the Ferrarese relations assuming other dark matter profiles
appear to be in disagreement (particularly that with an isothermal
profile, as noted above).

Fitting a function of the form $\mbh=A\mdh^{1.82}$ to the data in
Fig.~\ref{fig_bhm-dhm}, we estimate the zero-point of the $\mbh-\mdh$
relation in our redshift range to be $\mbh=10^{8.4\pm0.2}\,{\rm \msun}$ at
$\mdh=10^{12.5}\,{\rm \msun}$. This estimate was obtained through
minimising $\chi^{2}(A)$ for the fitted function, and the confidence
interval represents
$\Delta\chi^{2}=1$ limits. The exponent in the fitted model was chosen
as that from the Ferrarese relation assuming a Seljak profile. As a check
we repeated the analysis with the most extreme exponents from the
various other models discussed in this paper and found that all resulted in
a zero-point within the confidence interval quoted.
By comparison the Ferrarese relations at $z=0$
give a zero-point of $\mbh=10^{7.3}$, $10^{7.8}$ and
$10^{8.7}\,{\rm \msun}$
at the same dark matter halo mass for isothermal, NFW and Seljak
profiles respectively.  In agreement with the above,
the relation using the Seljak profile is the best match to the
data.  Given the uncertainties in the evolution of the $\mbh-\mdh$
relation, this is not direct evidence for or against a specific dark
matter halo profile.  We also note that our
zero-point for the  $\mbh-\mdh$ is consistent with that found for a
small sample of QSOs (but with larger dynamic range) by Adelberger \&
Steidel (2005).

% stuff from Lance and Ana
\subsection{Bias in the measured $\mbh-\mdh$ relation}

In addition to the random errors associated with the values of mean
black hole mass and dark halo mass that we deduce, we should also
consider whether there are any systematic biases that may arise in our
analysis of the $\mbh-\mdh$ relation.  Biases due to the uncertainties
in the estimates of $\mbh$ are discussed above.  However, there is one
further issue that is particularly important if we wish to compare
with relations such as those of Ferrarese (2002) at low redshift.

Ferrarese draws her objects from a sample of local galaxies with
measured black hole masses.  These are fairly evenly distributed over
a range of bulge (and hence inferred dark halo) masses.  On the other
hand, our QSO sample is drawn from the population of active galaxies,
selected by luminosity.  This tends to produce a Malmquist-type bias
towards larger $\mbh$.  If we consider that, at a given redshift,
objects will only be detected above a given $\mbh$ mass (neglecting for
the moment variation in Eddington ratio), then the
mass function of dark matter halos (with more halos at low mass) and
any scatter in $\mbh$ for a given $\mdh$ will cause an excess of
objects above the fiducial $\mbh-\mdh$ relation (i.e. with greater
$\mbh$).  This leads to a bias of the observed mean value of $\mbh$
above the true $\mbh-\mdh$ relation.  Allowing there to also be
scatter in the $L/\ledd$ relation is 
equivalent to moving the effective $\mbh$ limit, and only moves the
observed zero-point of the $\mbh-\mdh$ relation parallel to the true
relation.

The size of the Malmquist-type bias depends on both the steepness of the
mass function and the amount of intrinsic scatter in the relationship
between black hole and dark halo mass.  Both these quantities are 
rather uncertain, but we can estimate the possible size of the effect
in the following way.  First, we assume the same mass function of dark
halos that was assumed when deducing the clustering bias that led to
the inferred mean dark halo mass: namely a Sheth et al. (2001) mass
function with the cosmological parameters assumed in this paper.
Then, we assume that the black hole mass function may be generated from
the dark halo mass function by applying the Ferrarese relation, either
with no evolution or with WL05 evolution in the
mass relation, but with some scatter in that relation. Constructing a
Monte-Carlo simulation of the $\mbh-\mdh$ relation as outlined above,
and including a cutoff in black hole mass corresponding to the flux
limit of the 2QZ we can find the magnitude of this effect on our
results. Note that a cut in $\mbh$ will also
cause us to overestimate the gradient of the relation. However, due
primarily to the small dynamic range in Fig.\ref{fig_bhm-dhm}, we
make no attempt to estimate the slope of the $\mbh-\mdh$ relation
in this work. We are concerned only with the zero-point of the
relation, which is also biased by the cut.

We test three values of scatter in the $\mbh-\mdh$ relation.  The
minimum scatter we adopt is the scatter between
black hole mass and bulge velocity dispersion or luminosity
inferred by Marconi et al (2004), namely a factor of 2 in black hole
mass.  However, there is likely to be substantial 
additional scatter in the relation between bulge mass and dark halo
mass: from Ferrarese (2002) we estimate the total scatter could be a 
factor 4 in black hole mass although without direct measurements of
dark halo mass this is difficult to assess reliably.  Table
\ref{tab:bias} shows the mean black hole masses after correcting our
measured values for this Malmquist-type bias in each redshift bin for
the two different evolutionary models and three different value of the
scatter in the $\mbh-\mdh$ relation, namely $\times2$, $\times4$ and
$\times10$.  Although we estimate above that the scatter should be
between a factor of 2 and 4, the uncertainties are great enough that
it is worth considering the affect of a larger scatter
(i.e. $\times10$). 

\begin{table}
\caption{
Estimates of Malmquist bias in the mean $\mbh$ measured relative to
the true $\mbh-\mdh$ relation.  For each redshift bin at
$\overline{z}$, we give the measured $\mbh$ (column 2) and the
corrected values assuming the three values of intrinsic scatter about
the mass relation ($\times2$, $\times4$ and $\times10$), for each of
the two assumed values for evolution (columns 3 to 8).  $\mbh$ is
given as log$_{10}(M)$ in solar units.  The final row gives the
zero-point calculated as described in the text.  The error on these
zero-points are $\pm0.22$ dex.} 
\label{tab:bias}
\setlength{\tabcolsep}{3pt}
\begin{tabular}{cccccccc}
\hline \hline 
& & \multicolumn{3}{c}{no evolution} & \multicolumn{3}{c}{WL05 evolution} \\
$\overline{z}$ & zero & $\times2$ & $\times4$ & $\times10$ &
$\times2$ & $\times4$ &  $\times10$\\
& scatter &  scatter & scatter & scatter & scatter & scatter & scatter \\
\hline
 0.556 & 7.8 & 7.75 & 7.61 & 7.26 & 7.74 & 7.59 & 7.25 \\
 0.803 & 8.1 & 8.04 & 7.90 & 7.53 & 8.05 & 7.90 & 7.56 \\
 1.208 & 8.3 & 8.24 & 8.08 & 7.72 & 8.25 & 8.10 & 7.74 \\
 1.224 & 8.4 & 8.35 & 8.19 & 7.81 & 8.35 & 8.18 & 7.84 \\
 1.414 & 8.5 & 8.44 & 8.26 & 7.88 & 8.44 & 8.27 & 7.91 \\
 1.552 & 8.5 & 8.44 & 8.26 & 7.88 & 8.44 & 8.28 & 7.91 \\
 1.585 & 8.9 & 8.83 & 8.64 & 8.24 & 8.84 & 8.67 & 8.29 \\
 1.746 & 8.9 & 8.83 & 8.64 & 8.23 & 8.85 & 8.68 & 8.30 \\
 1.919 & 9.0 & 8.94 & 8.74 & 8.32 & 8.95 & 8.77 & 8.39 \\
 2.132 & 9.0 & 8.94 & 8.72 & 8.28 & 8.94 & 8.75 & 8.35 \\
 2.445 & 9.1 & 9.01 & 8.77 & 8.32 & 9.02 & 8.84 & 8.44 \\
\hline
 z-p  & 8.4  & 8.33 & 8.14 & 7.74 & 8.33 & 8.16 & 7.79 \\
\hline \hline
\end{tabular}
\end{table}

The biases for the two evolution assumptions are similar, with the WL05
evolution producing slightly less bias at high masses and high redshift, 
as in this case the associated dark halo mass is lower for a given
black hole mass (and so the mass function is flatter).
For the minimum scatter assumption, the bias is $\sim 0.1$~dex.
For $\times4$ scatter, however, the bias increase to a factor
$\sim0.25$~dex or greater.  If we now consider the
zero-point for the $\mbh-\mdh$ relation derived above, we see that any
Malmquist bias will push the true zero-point to lower $\mbh$.    

We note that the true scatter in the $\mbh-\mdh$ is very poorly
constrained.  If it were to be considerably higher than the above
values, then the Malmquist bias would also be higher.  A scatter of a
factor of 10 in $\mbh$ for a given $\mdh$ will produce a bias of
$\sim0.6$~dex in the mean $\mbh$ with respect to the true $\mbh-\mdh$
relation. This would make our data inconsistent with the
Robertson et al. (2005) model (Eq. \ref{eq:rob2}) and the
Ferrarese (2002) model assuming a Seljak (2002) density DMH profile,
while giving better agreement with the NFW profile model in
particular.  Hence a detailed comparison of the high-redshift relation
deduced here and the relation at lower redshifts found by Ferrarese
(2002) requires a better understanding of the amount of Malmquist bias
in the QSO measurements, and in particular of the amount of scatter on
the $\mbh-\mdh$ relation. 

\subsection{The evolution in $\mbh$ and $L/\ledd$}

Fig.~\ref{fig_z-bhm}a displays evidence for a trend in estimated black hole
mass with $z$ as we observe $\mbh$ for our sample to drop by an
order of magnitude between redshift 2.5 and 0.5. The correlation
between $\mbh$ and $z$ is significant at the 99\% level (via a
Spearman rank test; i.e. the probability that the null
hypothesis of no correlation is correct is $<1$\%) and the evolution
in $\mbh$ is best characterised by $\mbh\propto(1+z)^{3.9\pm 1.1}$.
However, due to the flux limits of the 2QZ we can not use this to
infer black hole mass evolution in the global QSO population (see
section~\ref{sec:res}).

In our sample the flux limit of the 2QZ and the luminosity evolution of
QSOs conspire to put $L^{*}$ at a similar apparent magnitude at every
redshift we sample.  The mean luminosity of our sample (i.e. the third
column of Table \ref{tab:res}) scales as $(1+z)^{4}$ while
$L^{*}$ scales as $(1+z)^{3}$ (Table \ref{tab:res} lists the break in
the QSO luminosity function, $M_{b_{j}}^*$ at the mean redshift of
each bin using the polynomial form of Croom et al. 2004), so
that the range of differences between $L$ and $L^{*}$ for our sample
is equivalent to 1 magnitude.  We also note that the space density of
the QSOs in our redshift bins changes by only a factor of 2.7
\cite{2qz14} over our redshift range.  However, the strong luminosity
dependence of the virial black hole mass 
estimators, combined with this evolution in $L^{*}$, make untangling
luminosity and mass evolution difficult.  Indeed the evolution we
observe in BH
mass is entirely due to the luminosity component of the virial mass
estimator as the velocity widths show no significant trend with
redshift.  In principle QSOs in our sample could have shown evolution
in their emission line FWHM to alter the evolution of $\mbh$, but
this is not seen. The best approach to investigate BH mass evolution
in the global QSO population would be to construct a sample with larger dynamic
range in magnitude \cite{rich05}. This would allow sources of the same
luminosity to be compared over a range of redshifts.

We therefore restrict our discussion of evolution to $L^{*}$
QSOs. Since the 2QZ samples a range of luminosities around $L^{*}$ at
all redshifts, we can create a subsample defined by a luminosity
interval around $L^{*}$ which is not affected by the flux limits of
the 2QZ. The range in magnitude of this sample is defined at the
bright end by the absolute magnitude of a QSO at low redshift with
an apparent magnitude of $\bj=18.25$, and at the faint end
by the absolute magnitude of a source at high redshift with $\bj=20.85$.
Note that at $z=0.3$ (the lowest redshift in our sample)
$\mb^{*}=-22.59$ (calculated using the polynomial evolution model of
Croom et al. 2004). This corresponds to an
apparent magnitude of $\bj=18.15$, brighter than the flux
limit of the 2QZ. Hence we define the boundaries of this sample by
the absolute magnitudes corresponding to $\bj=18.25$ at $z=0.556$ and
$\bj=20.85$ at $z=2.445$, i.e. at the mean redshift of the end
bins. Thus the data for the end bins are still slightly affected
by the magnitude limits of the 2QZ. This new sample is then described,
at all redshifts, as all QSOs with
\begin{equation}
-0.62 < \mb-\mb^{*}(z) < 0.75.
\end{equation}

\begin{figure}
\centering
\centerline{\psfig{file=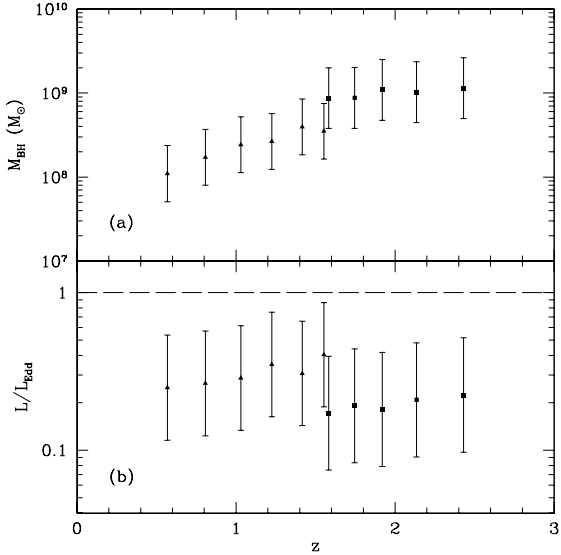,width=8.0cm}}
\caption{(\emph{a}) Black hole masses as a function of
  redshift for the sample of $L^*$ QSOs.
  (\emph{b}) Eddington ratios calculated for the same sample. The
  dashed line represents $L/\ledd=1$. In each case triangles and
  squares represent measurements from the \mgii\ and \civ\ lines 
  respectively.}
\label{fig_z-bhm_Lstar}
\end{figure}

We repeat the analysis described in this paper on this new sample to
investigate the evolution of black hole mass in typical $L^{*}$
QSOs. Composite spectra were made for the same redshift intervals, and
virial black hole masses were estimated from
these. Table~\ref{tab:res2} shows a summary of these results and they
are plotted in Fig.\ref{fig_z-bhm_Lstar}a. We
find $(1+z)^{3.3\pm1.1}$ evolution in $\mbh$, less
pronounced than in the whole sample although still marginally significant.
This shows that QSO samples at lower redshift are increasingly dominated
by lower mass BH, but as these are also lower-luminosity QSOs this cannot
directly be interpreted as evidence for anti-hierarchical "downsizing".

%This trend implies
%\emph{downsizing} of the active super-massive black hole population in the
%sense that the
%mass of typical $L^{*}$ QSOs declines towards low redshift. However,
%this is not the strict downsizing as observed in hard
%X-rays. In that case observations show lower-luminosity objects peak at lower
%redshift which implies, if we take luminosity to be a measure of
%$\mbh$, anti-hierarchical behaviour (low mass black holes grow faster
%at low $z$).

\begin{table}
\begin{center}
\caption{Summary of analysis on our sample of QSOs defined by a
  constant magnitude interval around $\mb^{*}$. Column~3 shows the
  difference between the average magnitude of the QSOs in that bin and
  $\mb^{*}$. Note that for all but the final redshift bin this is
  nearly constant, the final redshift bin (and the first) is affected
  by the magnitude limits of the 2QZ.}
\label{tab:res2}
\begin{tabular}{ccccc}
\hline \hline
$\overline{z}$ & $\overline{M}_{b_{\rm J}}$ & $\overline{M}_{b_{\rm
    J}} - \mb^{*}$ & Log$(\frac{\mbh}{\msun})$ & Log$(L/\ledd)$ \\
\hline
0.568 & -23.09 & 0.26 & $8.0\pm0.33$  & $-0.6\pm0.33$ \\
0.807 & -23.68 & 0.26 & $8.2\pm0.33$  & $-0.6\pm0.33$ \\
1.030 & -24.17 & 0.25 & $8.4\pm0.33$  & $-0.5\pm0.33$ \\
1.225 & -24.50 & 0.28 & $8.4\pm0.33$  & $-0.5\pm0.33$ \\
1.416 & -24.81 & 0.27 & $8.6\pm0.33$  & $-0.5\pm0.33$ \\
1.566 & -25.02 & 0.25 & $8.7\pm0.35$  & $-0.6\pm0.35$ \\
1.746 & -25.18 & 0.29 & $8.9\pm0.36$  & $-0.7\pm0.36$ \\
1.920 & -25.37 & 0.24 & $9.0\pm0.36$  & $-0.7\pm0.36$ \\
2.135 & -25.46 & 0.26 & $9.0\pm0.36$  & $-0.7\pm0.36$ \\
2.431 & -25.67 & 0.10 & $9.1\pm0.36$  & $-0.6\pm0.36$ \\
\hline \hline
\end{tabular}
\end{center}
\end{table}

To further elucidate what may drive QSO luminosity evolution we
calculate average Eddington ratios ($L/\ledd$) for the QSOs in our
redshift bins. In calculating the Eddington 
ratios bolometric luminosities ($L$) were found using the relation
derived by McLure \& Dunlop~(2004) for the $B$ band correcting
by $\bj=B-0.06$ for a mean QSO $B-V=0.22$ \cite{c+v90}. The relation is
then
\begin{equation} \label{equ_mbj}
\mb=-2.66\,{\rm log}(L)+79.42
%L=10^{\frac{79.42-\mb}{2.66}}
\end{equation}
for $L$ in watts. $\ledd$ (also in watts) is given by
\begin{equation} \label{equ_ledd}
\ledd=10^{39.1}\left(\frac{\mbh}{10^{8}\,{\rm \msun}}\right).
\end{equation}
McLure \& Dunlop arrive at this relation from the bolometric
corrections of Elvis et al. (1994). Elvis et al. find an error on their $B$
band bolometric correction of $\sim35$\,\%, and all of their
sample give a correction within a factor of $\sim2$ of the mean. Even
taking this factor of two as the error on our
bolometric correction for a single object, when
averaged over the $\sim2000$ QSOs in a redshift bin this translates to
a factor of only
$2/\sqrt{2000}\sim0.04$. This error is small when compared with
those on our black hole mass estimates, and is ignored in the
following analysis. Any luminosity dependence in the bolometric
correction would introduce systematics into our results, however,
Richards et al. (2006) find no evidence such a dependence.

Calculated Eddington ratios for the $L^*$ QSOs are plotted in
Fig.~\ref{fig_z-bhm_Lstar}b. We
observe no significant trend in Eddington ratio with redshift, characterised
by $L/\ledd\propto (1+z)^{-0.4\pm 1.1}$, indicating that the most
luminous QSOs that are typically observed in samples have similar Eddington
ratios, irrespective of redshift. Overall, however, the mean rate of
accretion in the universe must have been higher at high $z$, because the
integrated luminosity arising from AGN was higher at high $z$ than at low,
whereas the integrated irreducible mass in black holes cannot have been
greater at high $z$ than at low (see also Miller, Percival \& Croom
2005). It seems then
that the process of selecting luminous QSOs results in an almost invariant
Eddington ratio for the most extreme objects at any epoch, although given
both the statistical and systematic uncertainties present in
Fig.~\ref{fig_z-bhm_Lstar}b it is not yet possible to rule out some
redshift variation of Eddington ratio for these objects.

%Our calculated Eddington ratios for the sample of $L^*$ QSOs are plotted in
%Fig.~\ref{fig_z-bhm_Lstar}b.
%We observe no significant trend in Eddington ratios with redshift,
%characterised by $L/\ledd\propto(1+z)^{-0.4\pm 1.1}$.
%This indicates considerably less evolution in Eddington ratio than the
%theoretical value of $\sim(1+z)^{2.4}$ proposed by
%Miller, Percival \& Croom (2005). Our Eddington ratios, however, are
%for a sample of $L^*$ QSOs whereas they average over all
%galaxies in their calculations. Accordingly, our values for the
%Eddington ratios are also considerably higher than those given by
%Miller, Percival \& Croom (2005).

Our results are consistent with no evolution in Eddington
ratios over the redshift range studied. Instead we must conclude that
the luminosity evolution of $L^*$ QSOs is driven, at least for the most
part, by a reduction in black hole mass for $z<2.5$. This said the
confidence intervals on our evolution parameters are large and we
restrain from drawing any firm conclusions on evolution from these
data. The relative contribution of mass and Eddington
evolution are affected strongly in our data by the slope of the
radius-luminosity
relation used in the virial mass estimates
and the possibility of luminosity/redshift dependence in
velocity widths (which is small, see Corbett et al. 2003).
In addition the calibration of the virial mass estimators also has a
major affect on our results. These are not dependent on $z$ and hence
alterations to the calibrations would lead only to an offset in our
mass estimates. However, since each calibration applies to a
different range of redshifts, an offset to one of these calibrations
will significantly change our evolution results. We note that had we
performed the above analysis with older calibrations for the two mass
estimators we would have found considerably less evolution in $\mbh$ over this
redshift range, and a correspondingly larger evolution on Eddington ratios.

%We also note that (so far as we know) there is no redshift
%dependance in the virial mass estimators and we find that the
%emission line widths do not vary with redshift. Thus any future updates or
%recallibrations of the virial mass estimators will result only in an
%offset to the data presented here, and hence one might think that this
%would not affect the evolution measurments made. However, 

\section{Conclusions} \label{sec:conc}

We make composite spectra of QSOs from the 2QZ to find average virial
black hole mass estimates for the QSOs in 10 redshift bins for which
Croom  et al.~(2005) had already calculated $\mdh$ via clustering
analysis. Comparing the black hole and dark halo masses we
find evidence for $\sim(1+z)^{2.5\pm1.8}$ evolution in the $\mbh-\mdh$
relation,
%(i.e. $\mbh-\sigma$ constant)
although large errors are such that we can not exclude the possibility
of there being no evolution. We derive
the zero-point of the $\mbh-\mdh$ relation (averaged over redshift)
and find it to be $\mbh=10^{8.4\pm0.2}\,{\rm \msun}$ for a dark matter
halo of mass $\mdh=10^{12.5}\,{\rm \msun}$.  This is most consistent
with a model using a Seljak (2004) dark matter profile (under the
assumption of no evolution in $\mbh-\mdh$), however, uncertainties are
such that we are unable to definitively distinguish which model is
preferred.  We compare our measured $\mbh-\mdh$ relation to that
derived from hydrodynamical simulations of galaxy evolution
\cite{dsh05,rob05} and find good agreement.  

We note that because QSOs are selected above a given luminosity this
will tend to select objects above a given $\mbh$.  This results in a
Malmquist-type bias such that the observed mean $\mbh$ will lie above
the true $\mbh-\mdh$ relation.  The level of bias is crucially
dependent on the amount of scatter in the $\mbh-\mdh$ relation.

We take a subsample of QSOs in a constant magnitude interval
around $\mb^{*}$ and find significant evolution in their black
hole masses characterised by $\mbh\propto(1+z)^{3.3\pm 1.1}$.
Comparing this to the observed lack of significant evolution
in Eddington ratio ($L/\ledd\propto(1+z)^{-0.4\pm1.1}$) we conclude
that luminosity evolution of $L^*$ QSOs is driven primarily by
decreasing black hole masses between redshifts 2.5 and 0.5. However,
the exact combination of evolution in $\mbh$ and $L/\ledd$ is
dependent on the slope of the luminosity dependence in the virial mass
estimator and any luminosity/redshift dependence in the velocity width
of the QSO broad lines as well as the calibrations of the virial mass
estimators themselves. Considering this and potential sources of
systematic errors in our line width measurements, we find that our data are
still consistent
with a picture in which both reducing black hole masses and Eddington
ratios play an equal role in $L^*$ QSO luminosity evolution as observed in
other studies (e.g. Merloni 2004; Heckman et al. 2004).

To extend this work further, detailed analysis of samples with a
broader dynamic range will be required, including the analysis of QSO
clustering as a function of luminosity.  This has started to be done
with small samples (e.g. Adelberger \& Steidel 2005), but will be
extended with new faint QSO surveys such as the 2dF-SDSS LRG and QSO
(2SLAQ) Survey (Richards et al. 2005).

\section*{acknowledgements}

We warmly thank all the present and former staff of the
Anglo-Australian Observatory for their work in building and operating
both the 2dF and 6dF facilities.  The 2QZ and 6QZ are based on
observations made with the Anglo-Australian Telescope and the UK
Schmidt Telescope. 
We would also like to thank all of the good people at the University
of Sydney for their help, their advice and their on-going support.

\end{document}